\documentclass[%
reprint,
twocolumn,
groupedaddress,
amsmath,amssymb,
prl,
]{revtex4-1}
\usepackage{graphicx}
\usepackage{hhline}
\usepackage{multirow}
\usepackage{color}
\usepackage{times}
\usepackage{url}
\usepackage{tikz}
\usepackage{tikz-feynman}
\usepackage{dsfont}
\usepackage{mathtools}
\usepackage{dcolumn}
\usepackage{setspace}
\usepackage{ulem}
\usepackage{physics}
\usepackage{comment}
\usepackage{hyperref}
\usepackage{bbm}
\usepackage[bottom]{footmisc}

\usepackage[toc,page,titletoc]{appendix}
\usepackage{amsfonts}

\newcommand{\blue}[1]{ \textcolor{blue}{#1}}

\begin{document}
	\title{Quantum Valley Hall effect without Berry curvature}
	
	\author{Rasoul Ghadimi}
	\thanks{These authors contributed equally}	
	\author{Chiranjit Mondal}
	\thanks{These authors contributed equally}
	\author{Sunje Kim}
	\author{Bohm-Jung Yang}
	\email{bjyang@snu.ac.kr}

	\affiliation{Department of Physics and Astronomy, Seoul National University, Seoul 08826, Korea}
	\affiliation{Center for Theoretical Physics (CTP), Seoul National University, Seoul 08826, Korea}
	\affiliation{Institute of Applied Physics, Seoul National University, Seoul 08826, Korea}
	
	\date{\today}
	
	\begin{abstract}
		The quantum valley Hall effect (QVHE) is characterized by the valley Chern number (VCN) in a way that one-dimensional (1D) chiral metallic states are guaranteed to appear at the domain walls (DW) between two domains with opposite VCN for a given valley.
		Although in the case of QVHE, the total Berry curvature (BC) of the system is zero, the BC distributed locally around each valley makes the VCN well-defined as long as inter-valley scattering is negligible.
		Here, we propose a new type of valley-dependent topological phenomenon that occurs when the BC is strictly zero at each momentum. 
		Such zero Berry curvature (ZBC) QVHE is characterized by the valley Euler number (VEN) which is computed by integrating the Euler curvature around a given valley in two-dimensional (2D) systems with space-time inversion symmetry. 
		1D helical metallic states can be topologically protected at the DW between two domains with the opposite VENs when the DW configuration preserves either the mirror symmetry with respect to the DW or the combination of the DW space-time inversion and  chiral symmetries. 
		We establish the fundamental origin of ZBC-QVHE. 
		Also, by combining tight-binding model study and first-principles calculations, we propose stacked hexagonal bilayer lattices including h-BX (X=As, P) and large-angle twisted bilayer graphenes as candidate systems with robust helical DW states protected by VEN.
	\end{abstract}
	\date{\today}
	\maketitle
	
	{\it Introduction.|}
	Quantum valley Hall effect (QVHE) is a topological phenomenon arising from nonzero Berry curvature (BC) around two valleys in time-reversal $\mathcal{T}$ symmetric systems with broken inversion $\mathcal{P}$ symmetry~\cite{PhysRevLett.102.096801}.
	The valley Chern number (VCN) is defined as the integration of Berry curvature (BC) for a given valley, and its change over a domain counts the number of one-dimensional (1D) chiral metallic states at the domain wall (DW)~\cite{doi:10.1073/pnas.1308853110,PhysRevX.3.021018,PhysRevB.84.075418}.
	Although the total BC of the system is zero due to $\mathcal{T}$-symmetry,
	if the BC is well-localized near each valley and the intervalley scattering is negligible, QVHE is robust.
	Bernal-stacked bilayer graphene is a representative system that exhibits QVHE under vertical electric field breaking $\mathcal{P}$ symmetry~\cite{PhysRevLett.99.236809,PhysRevX.9.031021,PhysRevB.89.085429,PhysRevLett.100.036804,PhysRevLett.101.087204,doi:10.1021/nl201941f,PhysRevB.92.041404,Wang_2021,PhysRevLett.127.116402,Ghader2020,PhysRevB.87.155415,Kim2020,PhysRevB.99.115410}.
	Many interesting ideas have been proposed to realize QVHE not only in solid state systems~\cite{Ju2015,Li2016,Yin2016} but also in classical wave systems~\cite{PhysRevB.98.155138,Lu2017,Dong2017,PhysRevLett.120.063902, 10.1063/5.0127559, PhysRevLett.120.116802,PhysRevLett.120.116802}. These QVHE setups have potential application for valleytronics~\cite{PhysRevB.77.235406,PhysRevLett.124.037701,Mrudul_2021,PhysRevApplied.19.034056}.

	The BC is an essential ingredient in various topological phenomena including the QVHE \cite{PhysRevB.108.L121405,RevModPhys.95.011002,doi:10.1126/sciadv.aaq0194}. 
	However, the recent studies of symmetry-protected topological states have uncovered various intriguing topological states that exist even in the absence of the BC \cite{ PhysRevLett.118.056401,PhysRevLett.121.106403,PhysRevB.99.235125,PhysRevLett.125.053601,PhysRevB.104.085205,PhysRevB.103.205303,Zhao2022,PhysRevB.108.075129,jiang2022experimental,PhysRevB.108.125101,Park2021,PhysRevLett.119.226801}.
	One notable example is the topological states in two-dimensional (2D) systems with space-time inversion ($\mathcal{I_{\text{ST}}}$) symmetry, which appears in the form of either the combination of $\mathcal{T}$ and $\mathcal{P}$ in the absence of spin-orbit coupling or the combination of $\mathcal{T}$ and two-fold rotation about the $z$-axis normal to the 2D plane ($\mathcal{C}_{2z}$), irrespective of the presence or absence of spin-orbit coupling \cite{Ahn_2019}.
	Although $\mathcal{I_{\text{ST}}}$ symmetry forces the BC to vanish at every momentum, two isolated bands in 2D $\mathcal{I_{\text{ST}}}$ symmetric systems can be characterized by an integer Euler number $\chi$ [see Eq.~(\ref{equ:Eulerclass}) for its mathematical definition] \cite{PhysRevX.9.021013}, which induces various intriguing physical properties \cite{Bouhon2020}.

	For instance, for a pair of $I_{ST}$-symmetric bands with the Euler number $\chi$, there must be band crossing points in-between in a way that the total vorticity of band crossing points must be equal to $2\chi$~\cite{PhysRevX.9.021013}.
	Moreover, a topological phase transition changing the Euler number accompanies pair-creation and pair-annihilation of nodal points \cite{PhysRevX.9.021013} which can be characterized by non-Abelian braiding processes \cite{Peng2022,PhysRevB.105.085115,Bouhon2020,Qiu2023,Jiang2021,PhysRevB.102.115135,slager2022floquet,PhysRevB.106.235428,Guo2021}.
	Also, the Euler number is a fragile topological index characterizing the band topology of nearly flat bands in twisted bilayer graphene \cite{PhysRevX.9.021013,PhysRevB.100.195135,PhysRevB.99.195455,PhysRevLett.123.036401}, which reduces to a $Z_2$ second Stiefel-Whitney number when additional bands are included \cite{Ahn_2019}.
	However, contrary to the case of stable topological states, the topological Euler bands cannot host boundary in-gap modes unless additional symmetries are imposed to the system \cite{PhysRevX.9.021013, PhysRevLett.125.126403}, as is common in crystalline topological states~\cite{doi:10.1146/annurev-conmatphys-031214-014501}.
	
	In this work, we propose a new type of valley-dependent topological phenomenon appearing in the absence of the BC, dubbed a zero Berry curvature (ZBC) QVHE. The ZBC-QVHE is protected by the change of valley Euler number (VEN), similar to the conventional QVHE where the VCN changes over the domain wall. 
	A nonzero change of VEN can protect one-dimensional (1D) helical metallic states along the DW between two domains when the DW configuration (DWC) satisfies either a mirror symmetry about the DW or $C_{2z}$ together with a chiral symmetry.
	Here, the DWC indicates the structure composed of two domains having the opposite VEN with a DW in between.
	As a simple model to realize the ZBC-QVHE, we propose 2D bilayer structures such that each layer hosts QVHE in layer decoupled limit and the VCNs of the upper and lower layers have the opposite sign, thus the total system has zero BC at every momentum and no QVHE is expected.
	
	\begin{figure}[t!]
		\centering
		\includegraphics[width=0.85\linewidth]{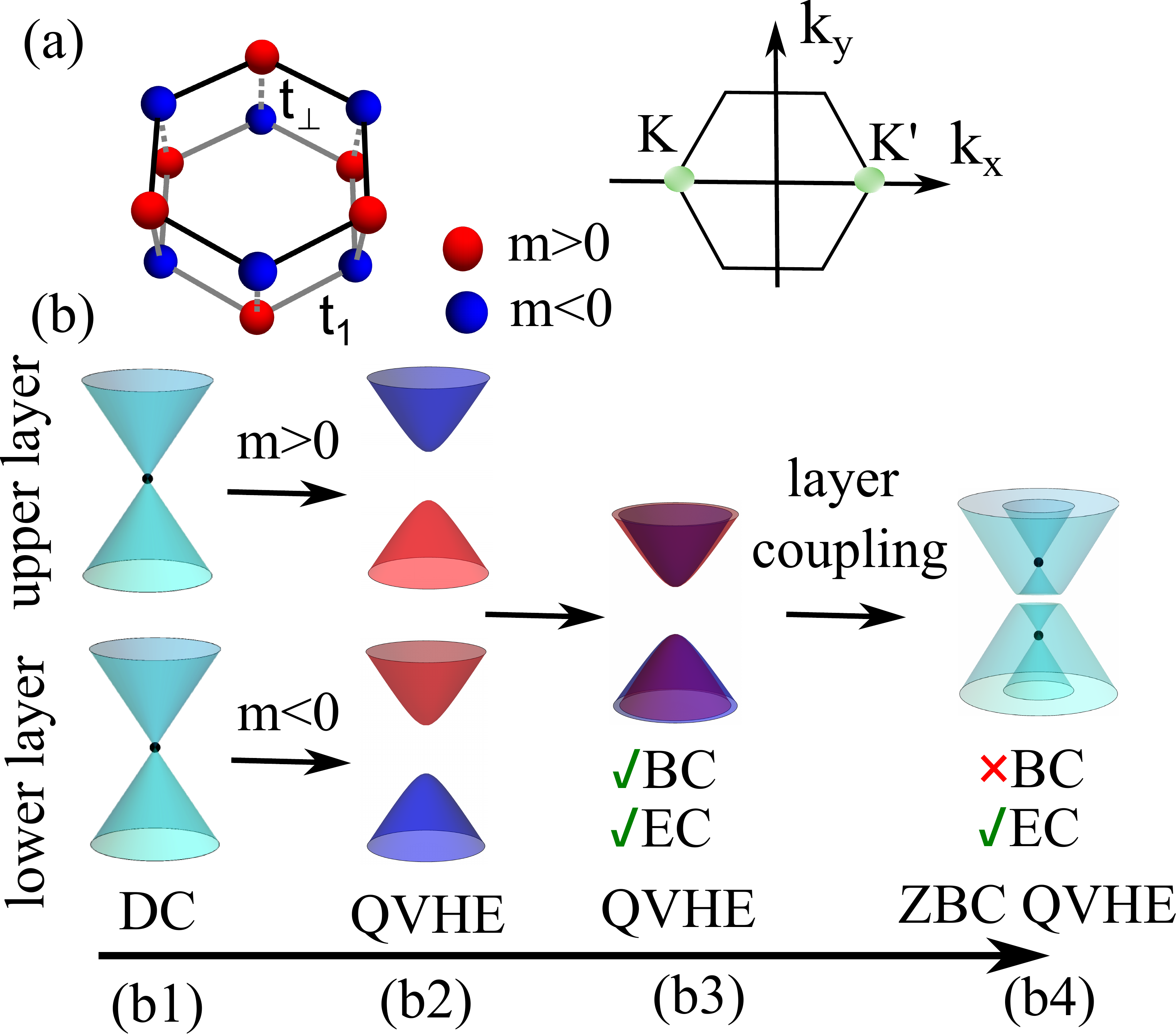}
		\caption{ 
			(a) Lattice structure of the AA'-stacked honeycomb bilayers and the relevant Brillouin zone with two valleys at $K$ and $K'$.
			(b) Schematic band structure evolution for zero Berry curvature quantum valley Hall effect (ZBC-QVHE). 
			(b1, b2) Two degenerate Dirac cones (DCs) are gapped out by oppositely signed masses, which gives oppositely signed Berry curvature (BC) (represented by blue and red colors, respectively) for the occupied bands. 
			(b3) Simple superposition of two decoupled gapped Dirac cones. 
			(b4) Coupling of two gapped Dirac cones, respecting space-time inversion symmetry. The BC vanishes but the Euler curvature (EC) is well-defined and nonzero.
		} 
		\label{fig1}
	\end{figure}


	{\it Model.|}
	To illustrate the idea, let us consider the {AA'}-stacked bilayer honeycomb lattice shown in Fig.~\ref{fig1}(a) with the nearest-neighbor hopping within each layer as well as between layers, and a staggered sublattice potential. 
	The low energy Hamiltonian at a valley (either K or K' as shown in Fig.~\ref{fig1}(a)) can be written as
	\begin{equation}\label{equ:doubleDirac}
		H_\eta(k_x,k_y)=v\eta k_x \sigma_x \tau_0+  v k_y \sigma_y \tau_0+ m\sigma_z \tau_z +t_{\perp}\sigma_0 \tau_x,
	\end{equation}
	where $v=\tfrac{2}{3} t_1$ is the Fermi velocity, $m$ is the staggered sublattice potential, and $t_{1}$, $t_{\perp}$ are the nearest neighbor intralayer and interlayer coupling, respectively. 
	$\eta=1, -1$ indicates $K$ and $K'$ valleys, respectively. $\sigma_i$ and $\tau_i$ $(i=x,y,z)$ are the Pauli matrices for the sublattice and layer degrees of freedom, respectively, and $\sigma_0$ and $\tau_0$ are corresponding $2\times 2$ identity matrices.

	We note that both $\mathcal{P}$ and $\mathcal{T}$ are symmetries of the full system but the Hamiltonian $H_\eta$ for each valley is invariant only under their combination $\mathcal{PT}\equiv \mathcal{I_{\text{ST}}}$, dubbed the space-time inversion symmetry, which can be represented by $\mathcal{I_{\text{ST}}}=\sigma_x\tau_x K$ with complex conjugation operator $K$ because $\mathcal{P}$ interchanges the layers and sublattices. 
	We note that $m$ is the only mass term allowed under $\mathcal{I_{\text{ST}}}$.
	When $m=0$, $H_\eta$ is invariant under two mirrors $M_{y}:y\rightarrow-y$ and $M_{z}:z\rightarrow-z$ represented by $M_y=\sigma_x$ and $M_z=\tau_x$, as well as two chiral symmetries $\Pi_0=\sigma_z\tau_z$ and $\Pi=\sigma_z\tau_y$.
	Here $\Pi_0$ is the bipartite chiral symmetry interchanging sublattices in the bilayer while $\Pi\propto M_z\Pi_0$.
	On the other hand, when $m\neq0$, $\{\Pi, H_\eta\}=0$ still holds while $M_y$, $M_z$, $\Pi_0$ symmetries are all broken.
	
	Fig.~\ref{fig1}(b) describes the band structure evolution of $H_{\eta}$ around a valley.
	When $m=t_{\perp}=0$, two degenerate Dirac cones, one from the top and the other from the bottom layers, appear at the Fermi energy ($E_F=0$)~[see Fig.~\ref{fig1}(b1)].
	Turning on nonzero $m$ generates oppositely signed masses at two Dirac cones, which results in opposite Berry curvature between them~[see Fig.~\ref{fig1}(b2)]. 
	Nonzero $t_\perp$ couples two gapped Dirac cones, which makes the BC vanishing at every momentum~[see Fig.~\ref{fig1}(b3, b4)].  
	However, both the upper and lower bands carry a half-integral Euler number, leading to ZBC-QVHE.
	Note that the presence of a single Dirac point in both the upper and lower bands (the blue dots in Fig.~\ref{fig1}(b4)) is also consistent with their half-integral Euler number \cite{Peng2022}.

\begin{figure}[t!]
		\centering
		\includegraphics[width=1\linewidth]{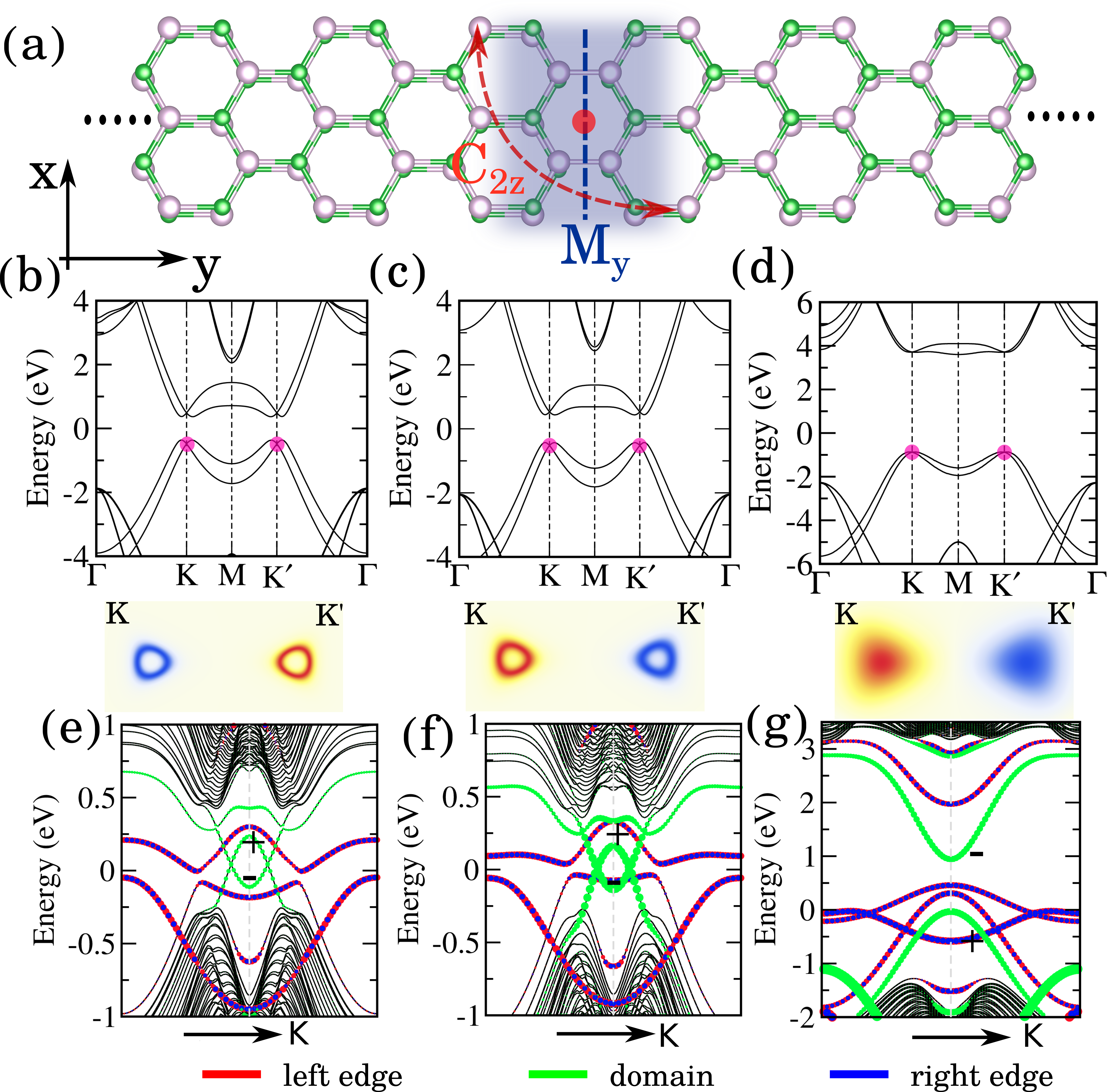}
		\caption{Realization of ZBC-QVHE in AA' stacked bilayer h-BX (X= As, P, N). 
			(a) Domain wall configuration (DWC) with $M_y$ and $C_{2z}$ symmetries. The blue dashed line and red disc represent the $M_y:y\rightarrow -y$ mirror and $C_{2z}$ rotation center, respectively.
			(b,c,d) Bulk band structure (top) and Euler curvature distribution (bottom) for (b)  h-BAs, (c) h-BP, and (d) h-BN, respectively. 
			$\mathcal{PT}$ symmetry-protected Dirac cones below the $E_F$ are indicated by the purple dots at  $K$ and $K'$ .
			(e,f,g)  DW modes (green lines) for h-BAs, h-BP, and h-BN, respectively, 
			where $\pm$ indicate $M_y$ mirror eigenvalues.
			As the bulk band gap increases (b to d), the Euler curvature spreads wider, and the crossing points between green lines get closer and eventually pair-annihilated (e to g). 
		}
		\label{fig2}
	\end{figure}

	{\it Topological invariant.|}
	When interlayer coupling is neglected with $t_{\perp}=0$, each layer carries a quantized VCN with the opposite sign between two layers.
	Explicitly, the valley Chern number $C_{\tau, \eta}$ for the valley $\eta$ in the layer $\tau$ is $C_{\tau, \eta}=\tfrac{1}{2}\text{sgn}(m\eta\tau)$ where $\tau=+1, -1$ indicate the top and bottom layers, respectively. 
	The change of VCN for each layer (and valley) over the domain wall $\Delta C^{v}_{\tau,\eta}\equiv 2 C_{\tau, \eta}=\text{sgn}(m\eta\tau)$ supports gapless DW modes between two domains for the given valley.
	Since the VCNs of two decoupled layers have the opposite signs, the DW of the bilayer shown in Fig.~\ref{fig2}(a) hosts two anti-propagating (helical) in-gap states per valley. 
	The existence of these helical in-gap states at the DW is nothing but the manifestation of the QVHE in each layer ~\cite{doi:10.1073/pnas.1308853110,PhysRevB.89.085429,PhysRevLett.100.036804,PhysRevLett.101.087204,doi:10.1021/nl201941f,PhysRevB.92.041404,Wang_2021,PhysRevLett.127.116402,Ghader2020,PhysRevB.87.155415,Kim2020}. 
	Interestingly, the helical DW modes survive even when $t_{\perp}\ne0$ and thus the total VCN vanishes. 
	Below, we show that the helical DW states are protected by the valley Euler number (VEN) of the bilayer.

	In $\mathcal{I_{\text{ST}}}$-symmetric systems, one can always find a basis such that both the Hamiltonian and wave functions becomes real \cite{PhysRevLett.118.056401,PhysRevX.9.021013,Bouhon2020}.
	In two dimensions, the non-abelian Berry curvature of two real bands, $\ket{u^1(\mathbf{k})}$ and $\ket{u^2(\mathbf{k})}$, forms a real antisymmetric matrix~\cite{PhysRevLett.118.056401,PhysRevX.9.021013}. Consequently, the usual Berry curvature, which is the trace of the non-abelian Berry curvature, vanishes, while its Pfaffian remains well-defined and gives the Euler curvature defined as $ \text{Ec}(\mathbf{k})  = \bra{\nabla u^1(\mathbf{k})} \times \ket{\nabla u^2(\mathbf{k})}$. By integrating the Euler curvature over the two-dimensional Brillouin zone (BZ), we obtain an integer topological invariant called the Euler number, given by
	\begin{equation}\label{equ:Eulerclass}
		\chi=\tfrac{1}{2\pi}\int_{BZ}\text{Ec}(\mathbf{k}) dk_xdk_y .
	\end{equation}
	We note that the above integral can also be performed on a disk $\mathcal{D}$ which is equal to  the patch Euler number if we neglect 1D integration along the boarder of $\mathcal{D}$~\cite{Bouhon2020,Peng2022}.
	We choose the size of $\mathcal{D}$ large enough to suppress the Euler curvature along the boundary, then the half-integer value of $\chi_{\mathcal{D}}$ can be determined by integrating the Euler curvature, similar to the valley Chern number calculation by integrating the BC around a valley. 
	For the coupled bilayer Hamiltonian in Eq.~(\ref{equ:doubleDirac}), the valley Euler number for a valley becomes $\chi_{\eta}=\tfrac{1}{2}\text{sgn}(m\eta)$, and the change of VEN over a DW is  $\Delta\chi_{\eta}^{v}\equiv 2\chi_{\eta}=\text{sgn}(m\eta)$.
	The DW between two domains with opposite VENs can host helical DW states when the system satisfies certain symmetry conditions, as discussed below.

	Note that by superposing two real bases, one can define the Chern bases $\ket{\psi^{\pm}(\mathbf{k})}\equiv(\ket{u^1(\mathbf{k})}\pm i \ket{u^2(\mathbf{k})})/\sqrt{2}$ such that the Chern number of $\ket{\psi^{+}(\mathbf{k})}$ ($\ket{\psi^{-}(\mathbf{k})}$)
	is equal (opposite) to the Euler number of real bases~\cite{Bouhon2020,PhysRevResearch.4.023188}. 
	When two layers are decoupled with $t_{\perp}=0$, the Chern bases are eigenstates of the Hamiltonian,
	which give the VCN of each layer. On the other hand, when $t_{\perp}\neq0$, the Chern bases are not eigenstates anymore, and thus only the VEN is well-defined.

	The relation between VCN and VEN is similar to that between the Chern number and $Z_2$ invariant in quantum spin Hall systems. 
	When two spin channels with the opposite Chern numbers are superposed, the Chern number of each spin channel is ill-defined if the spin-orbit coupling is present~\cite{PhysRevLett.95.226801,PhysRevLett.96.106802,PhysRevLett.95.136602} while their Chern number difference is well-defined modulo two when time-reversal symmetry exists~\cite{PhysRevLett.95.146802,PhysRevB.76.045302}.
	Similarly, in our case, the VCN of each layer is ill-defined when interlayer coupling presents, while the VEN is well-defined when $\mathcal{I_{\text{ST}}}$ symmetry exists

	{\it Stability of helical DW states.|}
	The helical DW states related to the VEN can be protected when the DWC
	satisfies either $M_y$ symmetry or the combination of chiral symmetry $\Pi$ and DW space-time inversion $\mathcal{I'_{\text{ST}}}\equiv\mathcal{C}_{2z}\mathcal{T}$ (see Fig.~\ref{fig2}(a)).
	The former ($M_y$ symmetry) is easily achievable and enough to protect the DW states while the latter also exists in various model Hamiltonians.
	Using the Hamiltonian in Eq.~(\ref{equ:doubleDirac}), 
	the DWC can be constructed by taking $m\rightarrow m(y)$, where $\lim_{|y|\rightarrow \infty}m(y)=\text{sign}(y) m_0$ and $m_0>0$. Accordingly, the DW is located along the $x$-direction. 
	The Hamiltonian for DWC can be obtained from Eq.~(\ref{equ:doubleDirac}) by substituting $k_y\rightarrow-i\partial_y$ as
	\begin{equation}\label{equ:domainwall}
		H^{\text{DW}}_\eta(k_x,y)=v\eta k_x \sigma_x \tau_0 -iv \partial_y \sigma_y \tau_0+ m(y)\sigma_z \tau_z +t_{\perp}\sigma_0 \tau_x.
	\end{equation}
	
	In the following, we show how $M_y=\sigma_x$ symmetry can protect the helical DW states.
	When the DWC satisfies $M_y$,
	the DW becomes a $M_y$-invariant 1D space. 
	Moreover, since $M_y$ changes the sign of $m$ in $H_{\eta}$, two domains related by $M_y$ should have the opposite Euler number.
	Similarly, when interlayer coupling is neglected, two domains in each layer related by $M_y$ should have the opposite  Chern number. 
	
	For convenience, let us first neglect interlayer coupling.
	Since each layer supports QVHE and the VCN of two layers have opposite signs, helical DW states should appear from each valley in the bilayer.
	The effective Hamiltonian for the helical DW states from one valley can be written as
	$H^{\text{DW}}_0=v'k_x\tau_z$ where $\tau_z$ gives a good quantum number when interlayer coupling is neglected.
	Since the DW is $M_y$ invariant, each DW state should also carry a $M_y$ eigenvalue.
	To have the $M_y$ eigenvalue and the layer quantum number simultaneously,
	$M_y$ should be represented by either $\tilde{M}_y=\tau_0$ or $\tilde{M}_y=\tau_z$ where the tilde symbol indicates the symmetry representation in the space spanned by helical DW states.
	$\tilde{M}_y=\tau_z$ prohibits the mass terms in the form of $H^{\text{DW}}_m=m_1\tau_x+m_2\tau_y$ even when interlayer coupling is turned on.
	
	The representation $\tilde{M}_y=\tau_z$ in the DW can be confirmed as follows. For simplicity, let us consider $H^{\text{DW}}(y,\tau)=-iv \partial_y \sigma_y + \tau m\sigma_z$ corresponding to Eq.~(\ref{equ:domainwall}) with $k_x=0$ and $t_{\perp}=0$, where $\tau=\pm1$ indicate the upper and lower layers, respectively.
	Because of the chiral symmetry   $\{H^{\text{DW}}(y,\tau),\sigma_x\}=0$, if $\psi(y)$ is a zero mode solution satisfying $H^{\text{DW}}(y,\tau)\psi(y)=0$, $\sigma_x\psi(y)$ is also a zero mode solution. Since there is only one zero mode per layer, two solutions should be proportional to each other as $\sigma_x\psi(y)=\lambda\psi(y)$ where $\lambda$ is a constant. Namely, $\psi(y)$ is an eigenstate of the chiral symmetry operator $\sigma_x$. 
	Moreover, as the mass term in $H^{\text{DW}}(y,\tau)$ has the opposite signs in two layers, the DW Hamiltonians for two layers are related as $\sigma_y H^{\text{DW}}(y,\tau=+1)\sigma_y=H^{\text{DW}}(y,\tau=-1)$. Thus, if $\psi(y)$ is a zero mode solution for one layer, $\sigma_y\psi(y)$ is the zero mode solution of the other layer. Since $\psi(y)$ and $\sigma_y\psi(y)$ have the opposite $\sigma_x$ eigenvalues and $\psi(-y)=\psi(y)$, $\tilde{M}_y=\tau_z$ can be confirmed 
	(see Supplemental Materials (SM) \footnote{See the Supplemental Materials (SM) for details on tight binding, derivation of edge modes, ZBC QVHE in twisted bilayer graphene, ZBC QVHE in canted antiferromagnetic systems, DW stability study, and DFT computation 
		\blue{which includes Refs. ~\cite{PhysRevB.108.L121405,PhysRevB.50.17953,PhysRevB.47.558,PhysRevB.59.1758,Mele_2012,PhysRevLett.101.056803,PhysRevLett.123.216803,PhysRevB.81.161405,PhysRevLett.114.226802,PhysRevB.106.035153,doi:10.1126/science.aar8412,PhysRevB.85.201105,PhysRevB.88.045429,PhysRevB.91.014202,PhysRevB.93.035123,PhysRevB.100.014510,PhysRevB.104.144511}.}} for additional discussion).
	This symmetry representation is not affected by interlayer coupling if it is small enough to maintain the band gap.
	Therefore, the DWC with $M_y$ symmetry can support helical DW states originating from VENs.
	Since the helical states have the opposite $M_y$ eigenvalues, their crossing is stable. 
	
	The helical DW states can also be symmetry-protected when $\Pi$ and $\mathcal{I}'_{\text{ST}}=C_{2z}T$ satisfying $[\Pi,\mathcal{I'_{\text{ST}}}]=0$ exist simultaneously. 
	As noted above, $\Pi$ is a combination of the bipartite chiral symmetry and $M_z$ mirror.
	Also, if the DWC has $C_{2z}$ symmetry, $\mathcal{I'_{\text{ST}}}=C_{2z}T$ exists and takes the form of $I'_{\text{ST}}=\sigma_x \mathcal{K}$ in Eq.~(\ref{equ:domainwall}).
	Since $\mathcal{I'_{\text{ST}}}$ changes the sign of $m$ in Eq.~(\ref{equ:doubleDirac}), two domains related by $\mathcal{I'_{\text{ST}}}$ have the opposite VENs.
	Since $C_{2z}$ does not mix layers, in the DW $\mathcal{I'_{\text{ST}}}$ can be represented by  $\tilde{\mathcal{I}}'_{\text{ST}}=K$ and forces $m_2=0$ in $H^{\text{DW}}_m=m_1\tau_x+m_2\tau_y$. 
	The remaining mass $m_1 \tau_x $ can be annihilated by imposing $\tilde{\Pi}=\tau_x$ that satisfies $[\tilde{\Pi},\tilde{\mathcal{I}}'_{\text{ST}}]=0$~\cite{PhysRevB.96.155105,PhysRevB.103.224523,PhysRevB.106.L121118}.

	{\it Hexagonal bilayer materials.|}
	As an example of AA' stacked hexagonal bilayers, we consider h-BX (B=Boron and X= As, P, N) as shown in Fig.~\ref{fig2}.
	Fig.\ref{fig2}(a) describes the DWC with $M_y$ and $C_{2z}$ symmetries composed of two domains with opposite VENs.
	The band structures of h-BX single domains and the corresponding Euler curvature distribution around two valleys are shown in Fig.\ref{fig2}(b,c,d) where we also identify the presence of a Dirac point between the two highest occupied bands at $K$ and $K'$ (purple dot in Fig.\ref{fig2}(b,c,d)), respectively, demonstrating their half-integral valley Euler number \cite{Peng2022}.  
	Our DFT simulation (see SM for detail about the computation method) of the DWC confirms the existence of helical DW modes in h-BAs, and hBP as shown in Fig.\ref{fig2}(e), and (f), respectively.
	Consistent with our prediction, the helical DW modes have opposite $M_y$ eigenvalues.

	In the case of h-BN, although the system respects all symmetry conditions to realize ZBC-QVHE, the helical DW modes are absent (see Fig.\ref{fig2}~(g)). 
	There are two reasons for this. 
	One is because, as the band gap increases (see Fig.~\ref{fig2} (b,c,d)), the wavefunctions at K and K' are mixed more, which makes VEN ill-defined.
	To confirm it, we have computed the Euler curvature distribution near two valleys for these systems shown in the bottom panels of Fig.~\ref{fig2} (b,c,d).
	One can clearly see that the Euler curvature spreads wider as the band gap increases.
	In h-BN with the largest band gap, the helical DW states from two valleys are pair-annihilated, leading to the gapped DW modes (see Fig.~\ref{fig2} (g)).
	
\begin{figure}[t!]
		\centering
		\includegraphics[width=0.9\linewidth]{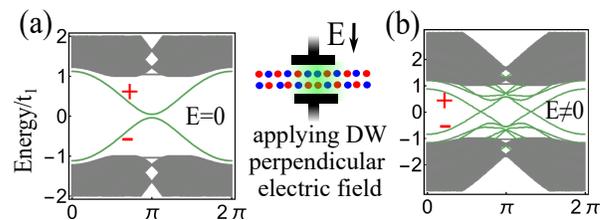}
		\caption{The evolution of the helical DW dispersion under an external vertical electric field which can restore the helical DW states. DW dispersion (a)  without and (b) with electric field.
		}
		\label{fig3}
	\end{figure}

	The other reason is due to the abrupt change of the on-site potential at the DW, roughly proportional to the sublattice potential, occurring due to the atomic configuration change as shown in Fig.~\ref{fig2} (a).
	As h-BN with the largest sublattice potential feels stronger potential variation at the DW, the helical DW modes from two valleys can be pair-annihilated. We note that as long as the DWC keeps the $M_y$ mirror, the crossing of helical DW states from each valley is stable even if such potential variation happens.
	Interestingly, this on-site potential variation at the DW can be compensated by applying an external electric field to the domain.
	As shown in Fig.~\ref{fig3}(a, b), the helical DW states can be recovered by applying an external electric field, which is obtained by using a tight-binding model relevant to h-BN with an additional out-of-plane electric field as described in detail in the SM.

	\textit{Discussion.|}
	We have proposed a new type of valley dependent topological phenomena, dubbed the ZBC-QVHE, induced by the VEN.
	Although the Euler number is a topological invariant of two bands, the helical DW states are stable even in the presence of additional bulk bands as long as they do not disturb the half-integral Euler number of Euler bands at each valley.
	This is because the crossing between helical DW states from each valley is further protected by $M_y$ symmetry of the DWC, which is well-defined even in multi-band systems.
	In the case of the helical DW states protected by $\tilde{\mathcal{I}}'_{\text{ST}}$ and $\tilde{\Pi}$ symmetries,
	the stability of the helical DW state crossing is guaranteed by the fact that $H_{\text{DW}}$ has non-zero $\mathcal{Z}_2$ $0$-dimensional charge when commuting $\tilde{\mathcal{I}}'_{\text{ST}}$ and $\tilde{\Pi}$ symmetries are present~\cite{PhysRevB.96.155105}. 
	
	In addition to h-BX materials, we have identified various candidate systems where ZBC-QVHE can be realized
	such as large-angle twisted bilayer graphene~\cite{PhysRevB.108.L121405,Mele_2012,PhysRevLett.101.056803,PhysRevB.81.161405,PhysRevLett.114.226802,PhysRevLett.123.216803,PhysRevB.106.035153}, 
	and hexagonal bilayer spin systems~\cite{Wang2019,ISLAM2016304,PhysRevLett.127.043904,doi:10.7566/JPSJ.84.121003,doi:10.1073/pnas.1219420110,PhysRevLett.109.206801}
	as described in detail in SM.  
	Furthermore, we anticipate that ZBC QVHE and its DW modes can be readily achieved in classical wave systems, where many QVHE phenomena have already been observed ~\cite{Lu2017, 10.1063/5.0127559, PhysRevLett.120.116802,PhysRevLett.120.116802,PhysRevLett.127.043904}.
	
	Finally, let us discuss the stability of the DW modes under various perturbations. 
	First, we examined the DW \(M_y\) symmetry-breaking effect by considering the unequal lengths of two domains as well as the vertical and lateral shifts of domains. 
	By performing first-principle calculations, we confirmed the stability of the DW modes against \(M_y\) symmetry breaking when the width of each domain is large enough and domain mismatch is within the experimentally accessible window~\cite{Note1}.
	Furthermore, we considered local disorder effects induced by random onsite potentials and local lattice distortions near the DW by analyzing finite-size lattices and performing the spectral function analysis~\cite{PhysRevB.85.201105,PhysRevB.88.045429,PhysRevB.91.014202,PhysRevB.93.035123,PhysRevB.100.014510,PhysRevB.104.144511}.
	As expected, in the presence of disorder, since the mirror symmetry is broken, the DW states lose their topological protection.  However, we found that even when the disorder potential strength is as large as the band gap, the only influence of local disorder effect is at most to open tiny gaps at the crossing of helical DW states. 
	Thus, our numerical study demonstrates the stability of the helical DW states against local perturbations, and their experimental feasibility~\cite{Note1}.

	\noindent \textbf{Acknowledgements}
	We thank Yuting Qian for the fruitful discussions. 
	R.G, C.M, S.K, and B-J.Y were supported by
	Samsung  Science and Technology Foundation under Project Number SSTF-BA2002-06 and
	the National Research Foundation of Korea (NRF) grant funded by the Korean government (MSIT) No. NRF-2021R1A5A1032996.
	This research was supported by GRDC (Global Research Development Center) Cooperative Hub Program through the National Research Foundation of Korea (NRF) funded by the Ministry of Science and ICT(MSIT) (RS-2023-00258359).

	\bibliography{refs}

\end{document}


	\title{SUPPLEMENTAL MATERIAL: Quantum Valley Hall effect without Berry curvature}
	
	\author{Rasoul Ghadimi}
        \author{Chiranjit Mondal}
        \author{Sunje Kim}
	\author{Bohm-Jung Yang}
	
	\date{\today}
 
 \maketitle
\tableofcontents
 
 \makeatletter
\renewcommand \thesection{S\@arabic\c@section}
\renewcommand\thetable{S\@arabic\c@table}
\renewcommand \thefigure{S\@arabic\c@figure}
\renewcommand{\theequation}{S.\arabic{equation}}
\makeatother

 \section{Tight binding and edge theory}
\begin{figure}[t!]
	\centering
	\includegraphics[width=0.7\textwidth]{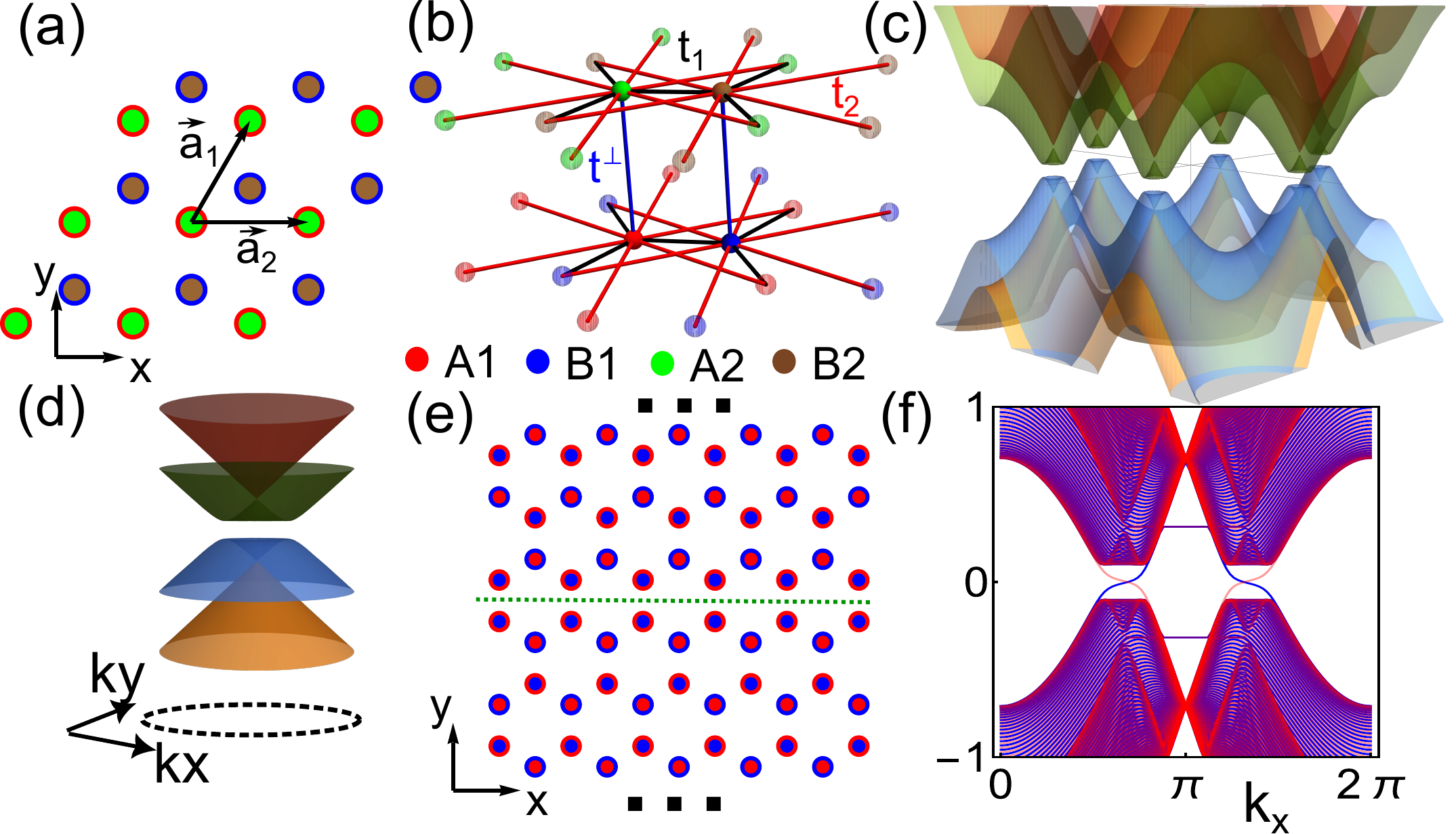}
	\caption{(a, b) Lattice structure and hopping parameter of  AA' stack of bilayer h-BX. 
 (c) The energy dispersion of Eq.~(\ref{equ:tightbinding}).
 (d) The magnified view of the (c) around a valley. The black dashed loop represents the boundary that we calculate the valley Euler number.
 (e) Domain wall structure, where the domain wall is represented by the dashed green line. The x direction is periodic. The red and blue disks show sites with opposite signs of onsite potential. 
 (f) Energy dispersion of domain wall of (e) is plotted [we set $t_{\perp}/t_1=0.3 $, $t_2=0$, and  $m/t_1=0.1$]. The red and blue colors in (f) indicated the given band's mirror eigenvalues ($\pm1$). }
	\label{fig:tightbinding}
\end{figure}

The simple tight-binding Hamiltonian of  AA'-stacked bilayer graphene can be read as [see Fig.~\ref{fig:tightbinding}(a, b)]
 \begin{equation}\label{equ:tightbinding}
 H=\sum_{\alpha,\alpha',l,l',i,j} (t^{\alpha, l;\alpha', l'}_{i,j} +m_{\alpha, l}\delta_{l,l'}\delta{\alpha,\alpha'}\delta_{i,j})\ket{\alpha,l,i}\bra{\alpha',l',j},
 \end{equation}
 where $\ket{\alpha,l,i}$, and $\bra{\alpha,l,i}$ are ket and bra states of  electron at given sublattice $\alpha=A, B$, layer $l=1,2$, and site $i$ index. In Eq.~(\ref{equ:tightbinding}) $t^{\alpha, l;\alpha', l'}_{i,j}$, and $m_{\alpha, l}$  indicate hoping  energy and onsite potential of electrons. 
 In the following, we consider $t_1$, and $t_2$ as the nearest-neighbor and next-nearest-neighbor coupling inside each layer, while the vertical hopping is given by $t_{\perp}$. 
 The staggered sublattice potential is considered as $m=m_{A, 1}=-m_{B, 1}=-m_{A, 2}=m_{B, 2}$, mimicking bilayer h-BX systems.  
 By applying the Fourier transformation, $\ket{\alpha,l,i}=\sum_{\mathbf{k}}\exp(i\mathbf{k}.\mathbf{r}_i) \ket{\alpha,l,\mathbf{k}}$, one can block diagonalize the Hamiltonian in the momentum space as $H=\sum_{\mathbf{k},\alpha,\alpha',l,l'} H_{\alpha,l;\alpha',l'}(k) \ket{\alpha,l,\mathbf{k}} \bra{\ket{\alpha',l',\mathbf{k}}} $, where $H(\mathbf{k})$ is a matrix given by
 \begin{equation}
     \label{equ:tightbindingmomentumspace}
     H(\mathbf{k})=\epsilon(\mathbf{k})\sigma_0\tau_0+h_x(\mathbf{k}) \sigma_x\tau_0+h_y(\mathbf{k}) \sigma_y\tau_0 + m \sigma_z\tau_z+ t_{\perp} \sigma_0\tau_x.
 \end{equation}
 $\sigma_i$ and $\tau_i$ $(i=x,y,z)$ represent the Pauli matrices for the sublattice and layer degrees of freedom, respectively, and $\sigma_0$ and $\tau_0$ are corresponding $2\times 2$ identity matrices. Here,   $\epsilon(\mathbf{k})$, $h_x(\mathbf{k})$, and $h_y(\mathbf{k})$ are function of momenta and given explicitly as 
\begin{equation}
    \epsilon(\mathbf{k})= 2 t_2 \left(2 \cos \left(\frac{\sqrt{3} k_x}{2}\right) \cos \left(\frac{3 k_y}{2}\right)+\cos \left(\sqrt{3} k_x\right)\right),
\end{equation}
\begin{equation}
    h_x(\mathbf{k}) = t_1 \left(2 \cos \left(\frac{\sqrt{3} k_x}{2}\right) \cos \left(\frac{k_x}{2}\right)+\cos (k_y)\right),
\end{equation}
\begin{equation}
    h_y(\mathbf{k}) =t_1 \left(\sin (k_y)-2 \cos \left(\frac{\sqrt{3} k_x}{2}\right) \sin \left(\frac{k_y}{2}\right)\right).
\end{equation}

 In the absence of $m, t_{\perp}$,   Eq.~(\ref{equ:tightbindingmomentumspace}) results in two degenerated Dirac cones at each valley $\eta=\pm1$, located at $K_{\eta}=\eta (\tfrac{4\pi}{3\sqrt{3}},0)$.
It is clear from  Fig.~\ref{fig:tightbinding}(e), that the domain wall is $\mathcal{M}_y$ symmetric. We plot the band structure for a ribbon that is periodic along the x-direction as shown in Fig.~\ref{fig:tightbinding}(f).
The mirror eigenvalue $+1$ and $-1$ of each band in Fig.~\ref{fig:tightbinding}(f) are indicated by red and blue colors.
We can see the helical states obtain opposite mirror eigenvalues, and therefore their crossing is stable. 
To prove this analytically, further, we derive domain wall states wave-function explicitly.
To start we first obtain the low energy model of Eq.~(\ref{equ:tightbindingmomentumspace}) by assuming small $m$ and $t_{\perp}$ at $K_{\eta}$  at a valley $K_{\eta}$
    \begin{equation}\label{equ:degQVHE}
    H=-3t_2 \sigma_0\tau_0+\tfrac{3}{2}k_x\eta t_1\sigma_x\tau_0+\tfrac{3}{2}k_yt_1\sigma_y \tau_0+ m\sigma_z\tau_z + t_{\perp} \sigma_0\tau_x. 
    \end{equation}
Note that nonzero $t_2$ only trivially shifts energy dispersion along the energy axis and therefore in the following discussion we set  $t_2=0$.  
    The domain is constructed by assuming $m\rightarrow m(y)$, where $\lim_{|y|\rightarrow \infty}m(y)=\text{sign}(y) m_0$.
    Accordingly, the domain is located along the $x$ direction. 
    The domain wall wave function is obtained by substituting $k_y\rightarrow-i\partial_y$. To proceed we furthermore assume  $k_x=0, t_{\perp}=0$ and later treat them perturbatively. 
    Therefore we need to solve
       \begin{equation}\label{equ:degQVHE1}
   -i\tfrac{3}{2}t_1\sigma_y \tau_0 \partial_y \ket{\psi(y)} + m(y)\sigma_z\tau_z \ket{\psi(y)}=0 
    \end{equation}
    where $\ket{\psi(y)}$ is a spinor wave function that depends on y and we assume that at $k_x=0$, the energy of domain modes is zero (see Fig.~\ref{fig:tightbinding}(f) that crossing is located at Dirac points). 
    To solve this equation we can express $\ket{\psi(y)}$ in a way that Eq.~\ref{equ:degQVHE1}  convert to a non-spinor equation. 
    This equation can be more simplified if we express it as $\ket{\psi(y)}=\sigma_z \ket{\psi'(y)}$,
     \begin{equation}\label{equ:degQVHE2}
   \tfrac{3}{2}t_1\sigma_x \tau_0 \partial_y \ket{\psi'(y)} + m(y)\tau_z \ket{\psi'(y)}=0 
    \end{equation}
By applying $\tau_z$ on both sides of equation we arrived at
    \begin{equation}\label{equ:degQVHE3}
   \tfrac{3}{2}t_1\sigma_x \tau_z \partial_y \ket{\psi'(y)} + m(y) \ket{\psi'(y)}=0 
    \end{equation}
    It is obvious that the $\ket{\psi'(y)}$ should be eigenstate of $\tau_z$ and  $\sigma_x$ or equivalently  $\ket{\psi'(y)}=f(y)\ket{\tau_z=\tau,\sigma_x=\sigma}$,     where  $\sigma=\pm1$ and $\tau=\pm1$. Here $f(y)$ is a y-dependent function that should be localized at the domain and can obtain by putting $\ket{\psi'(y)}$ in \ref{equ:degQVHE3},
    \begin{equation}\label{equ:degQVHE4}
   (\tfrac{3}{2}t_1\sigma \tau \partial_y f(y) + m(y) f(y))\ket{\tau_z=\tau,\sigma_x=\sigma}=0.
    \end{equation}
    Therefore to satisfying Eq.~\ref{equ:degQVHE4}, $f(y)$  should solve 
    \begin{equation}
         \tfrac{3}{2}t_1\sigma \tau \partial_y f(y) + m(y) f(y)=0.
    \end{equation}
 Now solving this equation is very easy and we obtain
    \begin{equation}
        f(y)=f(y_0)\exp[-\sigma \tau\int_{y_0}^y \tfrac{2}{3t_1} m(y')dy']
    \end{equation}
    Note that to describe the domain wall $f(y)$ should decay very fast to zero when $|y|$ increases. 
    Note that $\lim_{|y|\rightarrow \infty}m(y)=\text{sign}(y) m_0$. Therefore for large $|y|\gg 0$, we can obtain $f(y)$ as
    \begin{equation}
        f(y)\propto\exp[-\tfrac{2}{3t_1}\sigma \tau \text{sign}(y) m_0  y]
    \end{equation}
    or equivalently
        \begin{equation}
        f(y)\propto\exp[-\tfrac{2}{3t_1}\sigma \tau m_0 |y|]
    \end{equation}
    It is easy to see that $f(|y|\gg0)\rightarrow 0$ if  $-\sigma \tau \tfrac{2m_0}{3t_1}<0$ or equivalently $\sigma\tau=\text{sign}(t_1 m_0) $. This means that $\sigma$ and $\tau$ are not independent and therefore only two combinations of $\sigma=\pm1$, $\tau=\pm1$ can give localized domain wall states. 
    Combining the previous calculation, the following  ansatz gives a zero energy state at the domain wall 
\begin{equation}
    \ket{\psi^{\tau}_{\text{DW}}(y)}\equiv \tfrac{1}{\mathcal{N}}e^{-\text{sign}(m_0)\int_0^y \tfrac{2}{3|t_1|} m(y')dy' }  \sigma_z \ket{\tau_z=\tau, \sigma_x=\text{sign}(\tau m_0 t_1)},
\end{equation}
where we introduce $\mathcal{N}$ for normalization. 

The mirror symmetry is given by $\mathcal{M}_y=\sigma_x (y\rightarrow -y)$. 
Note that two domain wall states have opposite eigenvalues of $\sigma_x$. 
Therefore $\ket{\psi^{\pm}_{\text{DW}}(y)}$ are eigenstates of mirror symmetry and have opposite mirror eigenvalues.
Note that  mirror symmetry keeps y dependent function invariant, which can be proven as follows
\begin{equation}
    e^{-\text{sign}(m_0)\int_0^{-y} \tfrac{2}{3|t_1|} m(y')dy' }\underset{y'\rightarrow-y'}{=}
    e^{-\text{sign}(m_0)\int_0^{y} \tfrac{2}{3|t_1|} m(-y')d(-y')}\underset{m(-y')=-m(y')}{=}
    e^{-\text{sign}(m_0)\int_0^{y} \tfrac{2}{3|t_1|} m(y')d(y')}.
\end{equation}

To obtain domain wall Hamiltonian, we have to project other terms like $\tfrac{3}{2}k_x\eta t_1\sigma_x\tau_0+t_{\perp} \sigma_0\tau_x$ into zero energy domain wall modes. The first term gives 
$$H_{\text{DW}}\equiv \int_{\infty}^{\infty} dy \bra{\psi^{\tau}_{\text{DW}}(y)}\tfrac{3}{2}k_x\eta t_1\sigma_x\tau_0 \ket{\psi^{\tau'}_{\text{DW}}(y)} \propto  \tfrac{3}{2}|t_1|\text{sign}(\eta m_0) k_x (\tau_z)_{\tau,\tau'}.$$ 
But the second term $t_{\perp} \sigma_0\tau_x$ changes $\tau$  while keeping $\sigma$  intact. But as we saw before $\sigma$ and $\tau$ are locked and therefore this term gives zero if we project it to the domain wall states, equivalently
$$\int_{\infty}^{\infty} dy \bra{\psi^{\tau}_{\text{DW}}(y)}t_{\perp} \sigma_0\tau_x \ket{\psi^{\tau'}_{\text{DW}}(y)}=0.$$
This means that the domain wall crossing is protected even after coupling two layers.

To calculate the valley Euler number, we first assume the sublattice potential $m$ is small and then choose a patch around the Dirac point between two lower bands with a large radius to neglect the effect of $m$ on the border of the patch. 
Then we fixed the wave function on the boundary of the patch (black lines in Fig.~\ref{fig:tightbinding}(d)) and found that the sign of the valley Euler number changes with the change of the sign of sublattice potential $m$. 

\section{DFT computational details:}
First principle calculations were carried out using projector augmented wave (PAW)\cite{PhysRevB.50.17953} formalism based on Density Functional Theory (DFT) as implemented in the Vienna ab-initio Simulation Package (VASP).\cite{PhysRevB.47.558,PhysRevB.59.1758} 
The generalized-gradient approximation by Perdew-Burke- Ernzerhof (PBE)\cite{PhysRevB.59.1758} was employed to describe the exchange and correlation. 
An energy cut-off of 400 eV is used to truncate the plane-wave basis sets. 
DFT-D3 method has been applied for Van der Waals forces correction. 
We construct the domain of length around 180$\AA$ with at least 15$\AA$ vacuum size along the non-periodic directions to avoid any interactions between the periodic images.
The Brillouin zone (BZ) for the 1D ribbon is integrated using 1$\times$5$\times$1 $\Gamma-$centered k-mesh.
We relax the bilayer systems and then use the optimized lattice constants and layer separation distances to construct the 1D ribbon systems for h-BX (X=N, P, and As).
Due to the large size of the 1D ribbon, we only used the relaxed bulk parameters.
To calculate the Eular curvature of the h-BX, we construct a tight-binding Hamiltonian for the compounds using $p_z$ orbitals. With appropriate sublattice potential, we reproduce the band gap of the DFT results. 
Then we calculate the Eular curvature using the fitted Hamiltonian.

\newpage\section{Twisted bilayer graphene}
\begin{figure}[t!]
	\centering
	\includegraphics[width=0.7\linewidth]{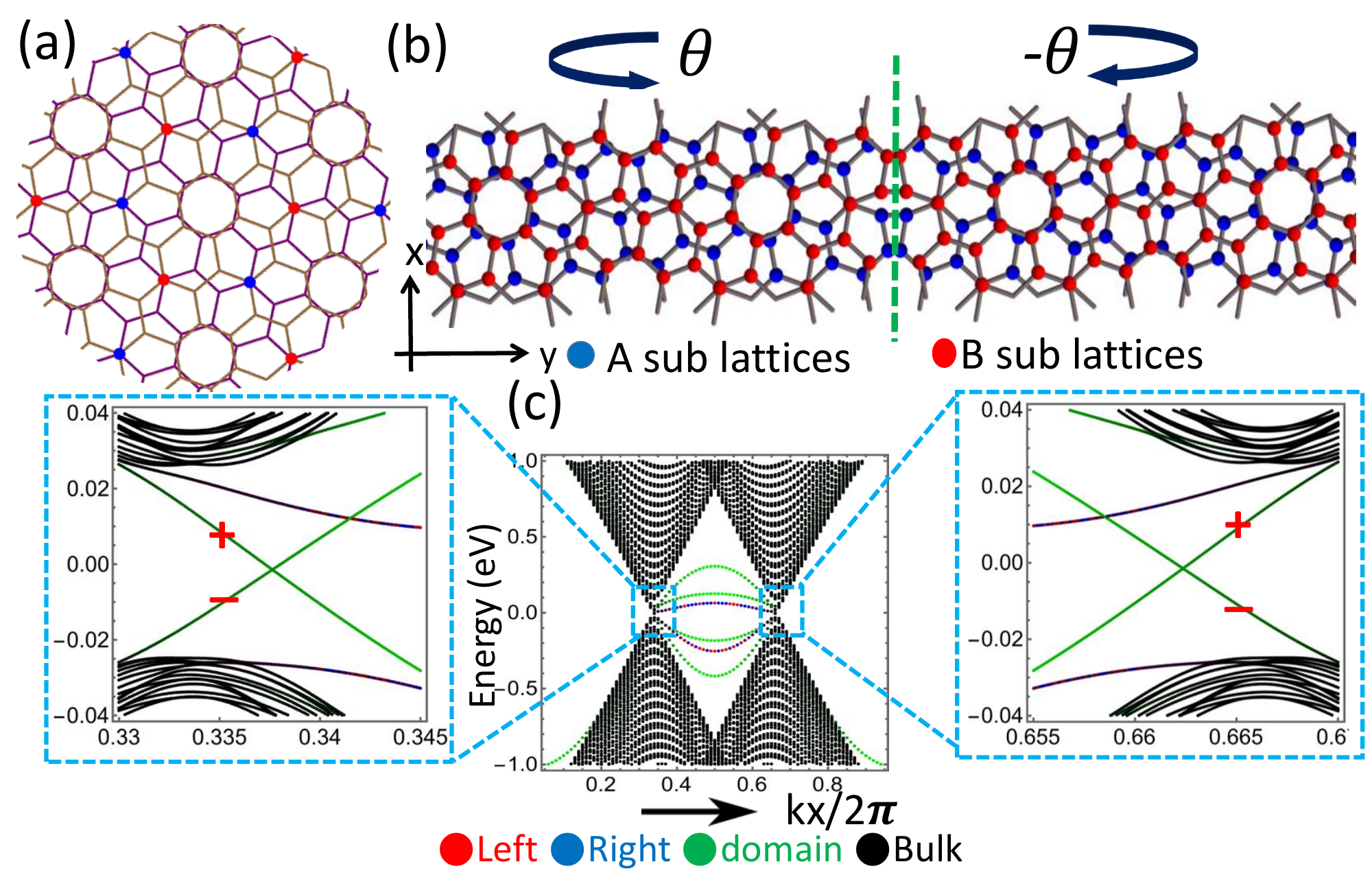}
	\caption{Realization of the ZBC QVHE in Twisted Bilayer Graphene (TBG). (a) The SEE configuration of TBG at $\theta=38.21$.  (b) The domain wall modes of ZBC QVHE can be achieved by symmetrically attaching two SEE configurations with opposite rotation angles $+\theta$ and $-\theta$. (c) Energy dispersion of the domain wall and its magnified versions around two valleys (we only show a few eigenvalues near the energy gap).  In Plots, the red plus and minus indicate opposite mirror eigenvalues for the given domain modes wave functions. }
	\label{fig3}
\end{figure}

In Ref.~\onlinecite{PhysRevB.108.L121405}, we investigated the QVHE in a large-angle TBG in its sublattice odd configuration ~\cite{Mele_2012,PhysRevLett.101.056803}. 
In the presence of a perpendicular electric field, this configuration acquires a non-zero valley Chern number and shows the QVHE.
However, the energy band dispersion of the sublattice even (SEE) configuration is the same as AA graphene but nodal lines are gapped out. 
The SEE configuration possesses $\mathcal{I_{\text{ST}}}$ as the combination of two-fold rotation and time reversal symmetry i.e., $\mathcal{C}_{2z} \mathcal{T}$ (Fig.~\ref{fig3}(a)) which leads to the absence of valley Chern number. It has been shown that the SEE configuration has a nontrivial Stiefel Whitney index, exhibiting corner modes ~\cite{PhysRevLett.123.216803}.
Therefore,  two occupied bands around the Fermi energy host non-trivial Euler class. 

The Hamiltonian of SEE configuration was derived by Mele in Ref.\cite{PhysRevB.81.161405} and is given by \cite{PhysRevLett.114.226802} 
\begin{equation}\label{equSM:TBG}
    H=\nu(k_x \eta \sigma_x+ k_y\sigma_y )+\gamma e^{i \phi \eta \tau_z \sigma_z} \tau _x e^{-i \phi \eta \tau_z \sigma_z},
\end{equation}
where $\sigma$, $\tau$, and $\eta$ are defined as sublattice, layer and valley degree of freedom.
Here $\gamma$ and $\phi$ are the parameters which depend on the rotation angle.
Here  $\mathcal{C}_{2z} \mathcal{T}=\sigma_x K$.
This Hamiltonian can be represented by 
\begin{equation}
    H=\nu( k_x \eta\sigma_x+ k_y\sigma_y )
    +\gamma \cos 2\phi \tau_x- \eta\gamma \sin 2\phi \tau_y\sigma_z.
\end{equation}
By transferring basis using a unitary matrix $U=e^{i\pi/4\tau_x}$ we obtain
\begin{equation}
    U^{\dagger}HU=\nu( \eta k_x\sigma_x+ k_y\sigma_y )
    +\gamma \cos 2\phi \tau_x+ \gamma \eta \sin 2\phi  \sigma_z\tau_z
\end{equation}
which is same as  the Hamiltonian for ZBC QVHE in the main text (Eq.~(1)), where mass and inter-layer coupling are replaced by $\gamma \eta \sin 2\phi$, and $\gamma \cos 2\phi $, respectively.  Therefore, changing the sign of $\phi$ changes the valley Euler number.
Therefore the sign of the gap can be determined by the sign of $\phi$.
As we can see easily from Eq.~\ref{equSM:TBG}, we can flip the sign of $\phi$ by  $\tau_x H \tau_x$.  Such operation is equivalent to the changing of layers, which is equivalent to twisting with the opposite angle. 
Furthermore sign of $\phi$ can be exchange using a mirror operator $\mathcal{M}= \sigma_x(k_y\rightarrow -k_y)$.
Also the sign of the $\phi$ can be exchange using a $P\mathcal{T}$ operation, $P\mathcal{T}= \sigma_x \tau_x K$.

To achieve the boundary modes related to ZBC QVHE, we designed a setup, that involves the joining of two twisted bilayers with opposite commensurate twisting angles and preparing the interface to respect the domain wall mirror symmetry (and/or domain wall space-time symmetry $I'_{ST}=\mathcal{P}\mathcal{T}$) as shown in Fig.~\ref{fig3}(b). 
We found that the energy dispersion of the system shows gapless helical states for each valley and it unequivocally confirms the ZBC QVHE in the SEE configuration of large-angle TBG (see Fig.~\ref{fig3}(c)). 
Furthermore, we anticipate the existence of such a ZBC QVHE in various other twisted bilayer systems, including $\alpha$-graphyne as $\alpha$-graphyne possess exactly similar band dispersion and contain the same symmetries ~\cite{PhysRevB.106.035153}.
Note that although in h-BX (X=N, P, and As) type stacked bilayer system, the ZBC QVHE can be linked easily to the existence of two physical QVHE, in twisted bilayer graphene it is difficult to obtain such a simplified but physical picture.

\newpage\section{Magnetic systems}

\begin{figure}
	\centering
	\includegraphics[width=0.7\linewidth]{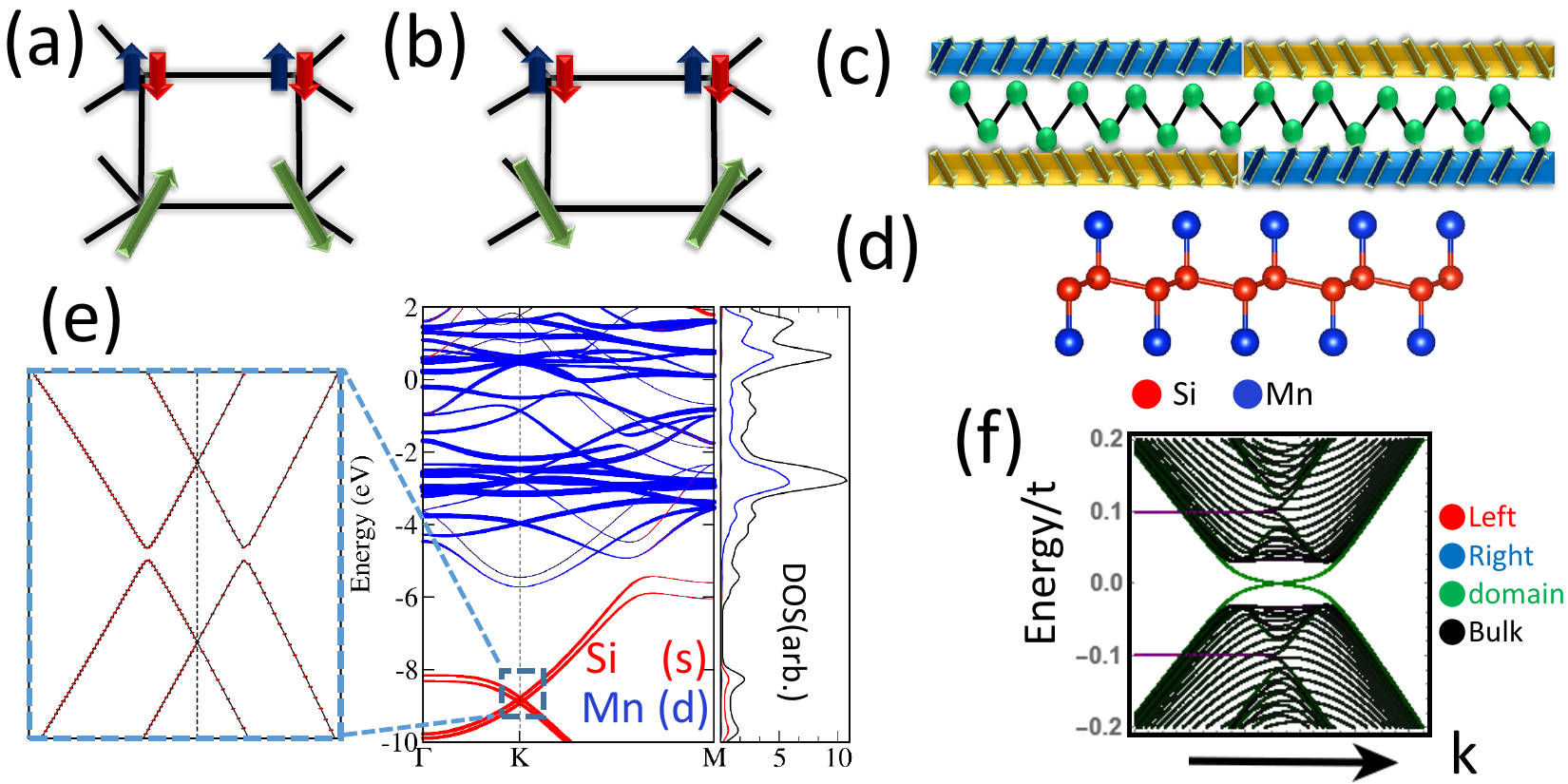}
	\caption{Realization of the ZBC QVHE in the anti-ferromagnetic and canted anti-ferromagnetic system.
 Degenerate magnetic moments with canted ground states are shown in (a) and (b).  
 (c) Proximity-induced canted anti-ferromagnetic states in buckled honeycomb layer and domain wall construction by controlling the magnetic proximity. 
 (d) Buckled SiMn monolayer.
 (e) Band structure of SiMn with canted magnetic ordering for Mn atoms.  
 (f)  The energy dispersion of electrons coupled to the canted magnetic domain structure shows ZBC QVHE.}
	\label{fig4}
\end{figure}

The layer degrees of freedom can be replicated using the spin sector of the magnetic system. 
The spin configurations in Fig.~\ref{fig4}(a,b) produce a similar Hamiltonian as the gapped bilayer AA graphene. However, the layer degree of freedom is replaced by the spin degree of freedom.
The canted magnetic configuration (see Fig.~\ref{fig4}(a,b)) leads to the following Hamiltonian: 
\begin{equation}\label{eq:magnetic}
    H_{\text{magnetic}}= k_x \eta \sigma_x s_0 + k_y \sigma_y s_0 + M^{\text{AFM}}   ~\sigma_z s_z+ M^{\text{FM}}\sigma_0 s_x
\end{equation}
where $M^{\text{AFM}}$ is the antiferromagnetic part and  $M^{\text{FM}}$ is the ferromagnetic part (which may arise due to an in-plane magnetic Zeeman field.). 
$\sigma_i$, and $s_i$ act on sublattice and spin degree of freedoms.  
The local  symmetry and chiral symmetry are given by  $\mathcal{I_{\text{ST}}}=\sigma_x s_x \mathcal{K}$ and $\Pi=\sigma_z s_x$. 
Here, $\mathcal{I_{\text{ST}}}$ is coming as a combination of spin-full time-reversal $\mathcal{\tau}=s_y\mathcal{K} (\mathbf k\rightarrow -\mathbf k)$ and $\mathcal{C}_{2z}= \sigma_x s_z(\mathbf k\rightarrow -\mathbf k)$ symmetries.
We construct a domain by flipping the sign of the $M^{\text{AFM}}$  over the two sides of the domain wall as shown in Fig.~\ref{fig4} (c). 
The domain have a mirror symmetry $M=\sigma_x (k_y\rightarrow -k_y)$ which interchange domains. 
Such a domain also respect the domain wall space-time inversion  symmetry which is expressed as  $\mathcal{I'_{\text{ST}}}=\sigma_x \mathcal{K}$.
Two sides of the domain wall possess magnetic configurations similar to Fig.~\ref{fig4} (a) and (b) respectively.
For $M^{\text{FM}} = 0$, the system becomes purely anti-ferromagnetic which leads to quantum spin-valley Hall effects.
Assuming that the system hosts these two degenerate magnetic solutions,  one can naturally expect similar to the Ising model the emergence of magnetic domains due to temporal fluctuations.
 Furthermore, to achieve a more controllable setup, one can harness desired magnetic states in a buckled honeycomb-like lattice such as silicene by proper magnetic ad-atoms adsorption (Fig.~\ref{fig4}(d)) or magnetic proximity effect (Fig.~\ref{fig4}(c)). 
We found that a canted magnetic order on buckled silicene potentially leads to the realization of ZBC QVHE. 
To confirm such a scenario we artificially construct SiMn lattice structure and assume canted magnetic order on Mn atom (Fig.~\ref{fig4}(d)).
In Fig.~\ref{fig4}(e), we present the band structure of SiMn, where we observe that the d-orbitals of the magnetic Mn atoms, situated around the Fermi energy, induce an effective ZBC QVHE phase for the s-orbitals of silicene well below the Fermi energy (see the magnified view of Fig.~\ref{fig4}(e)). 
Fig.~\ref{fig4}(f) confirms the existence of Helical domain states at the interface of two systems with magnetic configurations shown in Fig.~\ref{fig4}(a) and Fig.~\ref{fig4}(b), respectively.

\newpage\section{Stability of domain wall modes}

\begin{figure}
		\centering
		\includegraphics[width=0.6\linewidth]{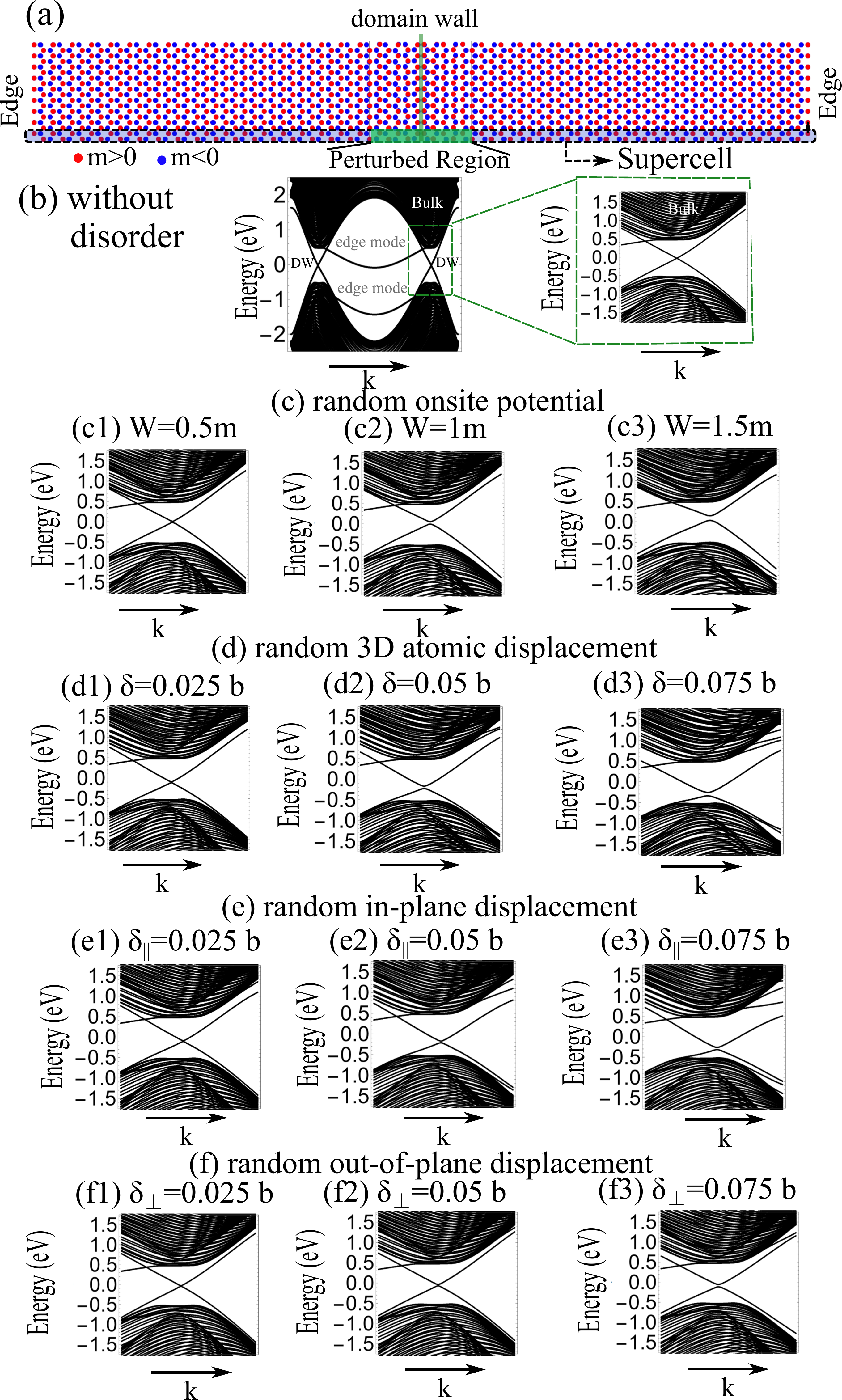}
		\caption{
			{Disorder effect with translational invariance along the domain wall (DW). 
				(a) Top view of DW. Blue and red sites indicate the atoms with opposite sublattice potentials.
                The bottom layer is behind and has the opposite color for vertices.
				The blue region at the bottom indicates the supercell.
				The green area shows the disordered region within a supercell ($2 \times 21$ atoms). 
				We assume translational symmetry along the vertical direction; therefore, the momentum $k$ and energy dispersions along the DW are well-defined.
				(b) The energy spectrum of the Hamiltonian in Eq.~(\ref{EQ1B}) with sublattice potential $m=0.5$ eV in the absence of disorder effect. Here, the supercell contains $2 \times 162$ atoms. The right panel shows a magnified view around a valley showing gapless helical modes.
				(c1-c3) The effect of random onsite potential $V_i \in [-W,W]$ with the strength \(W\) on the helical DW states.
				We note that even large disorders only induce a small gap in helical DW modes.
				In (c1-c3), we use the same random configurations and only increase the amplitude of $V_i$.
				(d-f) The effect of atomic distortions on the helical DW modes, where the atoms near DW are randomly displaced as $\vec{r_i} \rightarrow \vec{r_i} + (x_i,y_i,z_i)$ [in the green region in (a)]. We consider three different types of local distortions. 
				In (d), atoms are distorted in three dimensions, where $x_i,y_i,z_i \in [-\delta,\delta]$, and $\delta$ is the distortion strength. 
				We observe in (d1), that small distortion $\delta\ll b$ only induces a negligible gap, while   (d2) moderate and (d3) large distortions up to $\delta\approx 0.1 b$ ($10\%$) only induce a small gap in the DW helical modes.
				In (d1-d3), we use the same random distortion configuration and only increase the amplitude of $x_i,y_i,z_i$. 
				In (e1-e3), we assume the substrate effect is strong and does not allow the atom to move out of the plane and therefore assume $z=0, x_i,y_i \in [-\delta_{\parallel},\delta_{\parallel}]$. Similar to three-dimensional distortion in (d), even large distortion induces only a small gap in the DW helical modes. 
				In (e1-e3), we use the same random distortion and only increase the amplitude of $x_i,y_i$.
				In (f1-f3), for completeness, we also study out of the plane distortions and assume $x_i=0,y_i=0, z_i \in [-\delta_{\perp},\delta_{\perp}]$. 
				In (f1-f3), we use the same random distortion configuration and only increase the amplitude of $z_i$.
				As perturbation only induces a small gap in DW helical modes, we expect that the transport property of DW is not significantly affected. }
		}
		\label{fig1B}
	\end{figure}

	{In this section, We examined the stability of the DW modes under various disorders and lattice distortions using tight-binding model calculations and DFT simulations. 
	To construct the tight-binding model that simulates the AA' stacked h-BX material, we use a Slater-Koster tight-binding model that is used for graphene to describe the hopping energy, including the additional sublattice potential. More explicitly, the Hamiltonian is given as follows (we ignore the spin degrees of freedom):
	\begin{equation}\label{EQ1B}
		H = \sum_{i,j} t(\vec{r}_i - \vec{r}_j) c^\dagger_i c_j + \sum_i m_i c^\dagger_i c_i - \mu \sum_i c^\dagger_i c_i - \sum_i V_i c^\dagger_i c_i,
	\end{equation}
	where $\vec{r}_i = (x, y, z)$ is the position of the $i$-th site, $\mu \approx 0.8 ~\text{eV}$ is the chemical potential,  $m_i\in \pm m$ is the sublattice potential, $V_i$ is the onsite potential, and $t$ is the hopping term given by
	\begin{equation}
		t(\vec{r}) = \gamma_1 \frac{z^2}{r^2} e^{2.218(h_z - |r|)} - \gamma_0 \left(1 - \frac{z^2}{r^2}\right) e^{2.218(b - |r|)},
	\end{equation}
	in which $h_z = 3.349\, \text{\AA}$ and $b = 1.418\, \text{\AA}$ are the interlayer distance and the shortest carbon-carbon distance, respectively. The Slater-Koster parameters are given by $\gamma_0 = 2.7 \, \text{eV}$ and $\gamma_1 = 0.48 \, \text{eV}$ \cite{doi:10.1126/science.aar8412}.
	We assume the sublattice potential changes its sign over DW [see Fig.~\ref{fig1B}(a)].}

	{At first, without the disorder, the system has translational symmetry along the DW (the horizontal blue region at the bottom of Fig.~\ref{fig1B}(a) indicates a supercell that contains  $2 \times 162$ atoms), and therefore we can diagonalize Eq.~\ref{EQ1B} using a Fourier basis. We found that this model with $m = 0.5$ eV and domain wall configuration [Fig.~\ref{fig1B}(a)] shows helical DW modes per each valley as in the case of the h-BX system [see Fig.~\ref{fig1B}(b) for energy dispersion in the absence of disorder].  }

	{Now, let us consider the disorder effect which generally breaks translational symmetry along the DW.  
	As a source of disorder, we consider onsite potential fluctuations as well as local distortion of lattices around the DW that induces random hopping amplitudes.
	We treat the disorder effect in two different ways. One is to include the disorder effect in a finite region of a supercell near the DW while keeping the translational invariance along the DW. The other is to fully include the disorder effect, destroying the translational invariance along the DW.} 

	{Let us first discuss the former case. 
	As shown in Fig.~\ref{fig1B}, we introduce disorder in a finite region of length $8b$ (including $2 \times 21$ atoms) near the DW in a supercell [green perturbed region in Fig.~\ref{fig1B}(a)] while keeping the translational invariance along the DW. 
	This allows us to obtain the energy dispersion as a function of the momentum along the DW even in the presence of the disorder effect. }

	{In Fig.~\ref{fig1B}(c), we study the influence of the random onsite potential by assigning a random number $V_i \in [-W, W]$ with uniform distribution at the vertices inside the green perturbed region in Fig.~\ref{fig1B}(a).
	This type of random onsite potential may be induced by either the substrate or irregular atomic configurations near the DW.
	The energy dispersion of the DW states under weak disorder $W\approx 0.5m$ is shown in Fig.~\ref{fig1B}(c1) which is very similar to the disorder-free dispersion in Fig.~\ref{fig1B}(b) indicating the negligible disorder effect. 
	By increasing the disorder strength from the moderate level $W\approx m$ [Fig.~\ref{fig1B}(c2)] to large strength $W\approx 2m$ [Fig.~\ref{fig1B}(c3)], one can observe the opening of a tiny gap in the helical modes. Considering the small gap size, we expect that its influence on the charge transport property would be minor, especially if the chemical potential is located away from the gapped region.}

	{Now, let us discuss the influence of random hopping processes induced by local atomic distortions near the DW (green region depicted in Fig.~\ref{fig1B}(a)). 
	We consider three different types of atomic distortions: (i) atoms move in random 3D directions at the interface  [Fig.~\ref{fig1B}(d)], (ii) atoms move randomly in the in-plane directions at the interface [Fig.~\ref{fig1B}(e)], and (iii) atoms move randomly in the out-of-plane direction at the interface [Fig.~\ref{fig1B}(f)].}

	{Explicitly, in Fig.~\ref{fig1B}(d), we change the positions of atoms by $\vec{r}_i \rightarrow \vec{r}_i + (x_i, y_i, z_i)$ where the position fluctuations are described by $x_i, y_i, z_i \in [-\delta, \delta]$ with the distortion strength $\delta$. 
	Similar to the case of the onsite disorder, weak distortion has negligible influence on the dispersion of the helical DW modes [Fig.~\ref{fig1B}(d1)]. Even in the presence of large distortion, up to $10\%$ of the lattice constant, we found that the distortion induces a tiny gap [see Fig.~\ref{fig1B}(d2-d3)]. }

	{In Fig.~\ref{fig1B}(e), we consider the fluctuations of the in-plane positions $\vec{r}_i \rightarrow \vec{r}_i + (x_i, y_i, 0)$ where $x_i, y_i \in [-\delta_{\parallel}, \delta_{\parallel}]$ with the in-plane distortion strength $\delta_{\parallel}$. Similar to the case of the 3D random distortion, the influence of local distortion on the helical DW states is tiny even for large distortion up to $\approx 10\%$ of the lattice constant [see Fig.~\ref{fig1B}(e1-e3)]. }

	{In Fig.~\ref{fig1B}(f), we consider the out-of-plane positional fluctuations $\vec{r}_i \rightarrow \vec{r}_i + (0, 0, z_i)$ where $z_i \in [-\delta_{\perp}, \delta_{\perp}]$ with the vertical distortion strength $\delta_{\perp}$. Generally, consistent with the other cases discussed above, we found that even in the presence of large vertical distortion, the induced gap in the DW states is tiny.  }

	{To summarize up to this point, when we consider the disorder effect, including both onsite potential fluctuations and the lattice distortion (random hoppings), within a finite region of a supercell while keeping the translational invariance along the DW, we found that the helical DW states remain robust against weak to moderate disorder strength. The main influence of the disorder effect is to open a tiny gap in the energy dispersion of the helical DW states, whose effect on the transport properties is expected to be minor. }
	%

	\begin{figure}[t]
		\centering
		\includegraphics[width=0.7\textwidth]{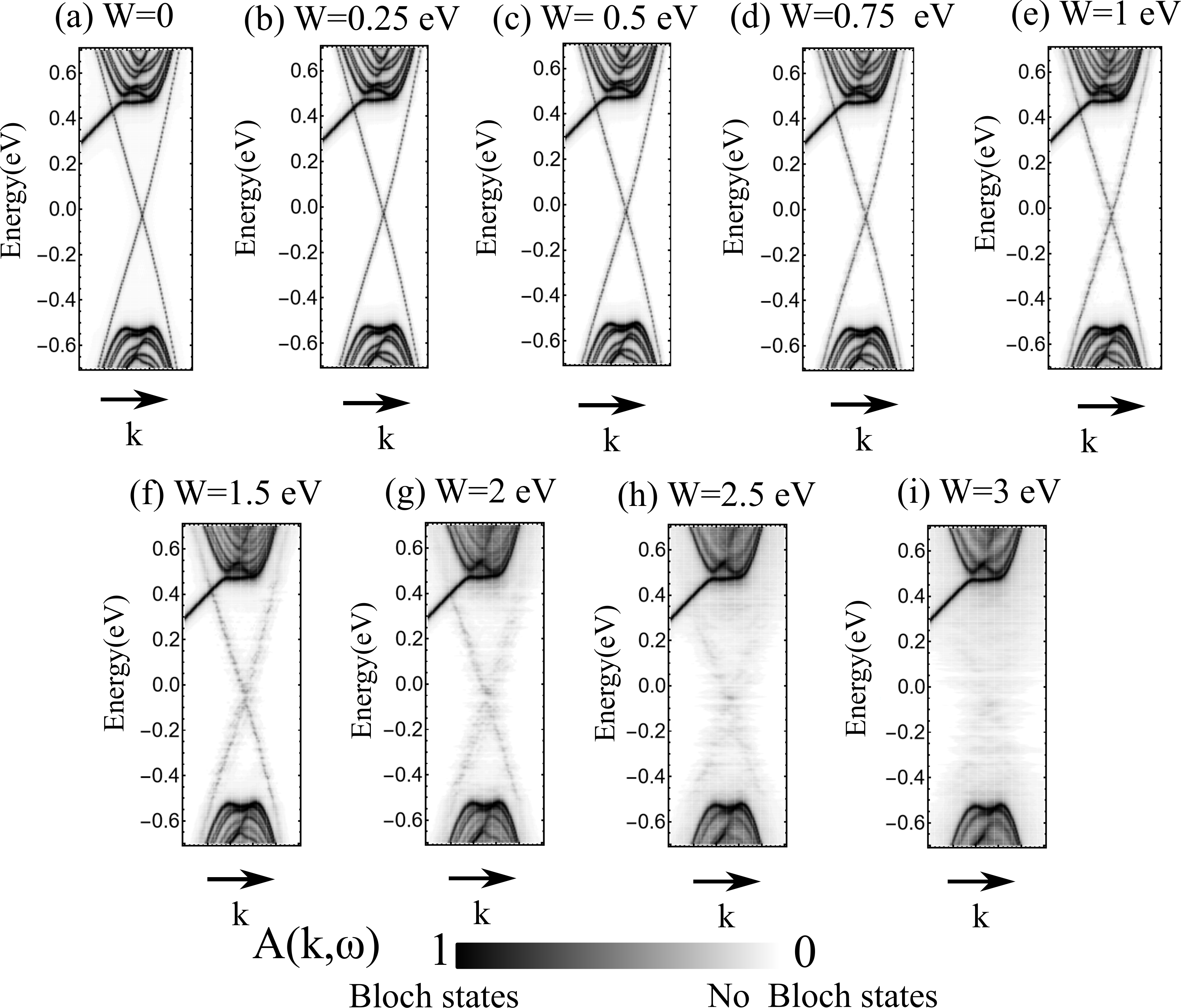}
		\caption{
			{Disorder effect in a finite-size system without translational invariance along the DW. 
				The spectral function $ A(k, \omega)$ (which shows the weight of DW Bloch states at given energy $\omega$ and momentum $k$) is plotted for several random onsite potential strengths $V_i\in [-W, W]$. 
				We introduce disorder only at the sites within the width of 8b near the DW which contains $2 \times 21$ atoms.  
				We solve the Hamiltonian in Eq.~(\ref{EQ1B}) with sublattice potential $m=0.5$ eV for a finite-size system composed of 285 supercells. Here, a supercell is composed of $2 \times 162$ atoms.
				In the plot, the opacity of lines shows the weight of the DW Bloch states in the spectral function. 
				The gapless helical states remain robust even with the disorder strength comparable to the gap $W\approx 2m$. Note that the sharpness of energy bands and the weight of the spectral function decrease by increasing disorder strength (for $W=3$ eV, for example, 80$\%$ reduction of the spectral weight appears compared to disorder-free case).  The weight of the spectral function becomes almost zero under strong disorder $W>\gamma_0$, where $\gamma_0=2.7$ eV is the Slater-Koster parameter that controls the bandwidth. In each panel, the random disorder strengths are given by (a) $W = 0$, (b) $W = 0.25$ eV, (c) $W = 0.5$ eV, (d) $W =0.75$ eV, (e) $W =1$ eV, (f) $W =1.5$ eV, (g) $W =2$ eV, (h) $W =2.5$ eV, and (i) $W = 3$ eV, respectively. }
		}
		\label{fig2B}
	\end{figure}

	{Now, let us consider the local disorder effect while completely ignoring the translational invariance along the DW.
	In this case, the momentum along the DW direction is not a good quantum number anymore. 
	However, by using the spectral function $A(k, \omega)$, one can study the effective DW energy dispersion in the presence of local disorder. }

	{More explicitly, to define the spectral function in the absence of translational symmetry, we need to directly diagonalize Eq.~\ref{EQ1B} for a finite system with open boundaries and obtain the eigenvectors $\ket{E_n}$ and eigenvalues $E_n$.
	The eigenvectors are given by $\ket{E_n}=\sum_i c_i^n \ket{i}$, where $c_i^n$ are numerically obtained coefficients indicating the wave function amplitude at the $i$-th atomic site.
	Then, the spectral function $A(k, \omega)$ is defined as
	\begin{equation}
		A(k, \omega) = \sum_{n,s} \delta(\omega - E_n) |\braket{s, k}{E_n}|^2,
	\end{equation}
	where $\ket{s, k}$ indicates the DW Bloch function given by $\ket{s, k} = \frac{1}{\mathcal{N}} \sum_{i \in s} \exp(i k \cdot x_i) \ket{i}$ in which $x_i$ is the $x$-position of the state $\ket{i}$ along the DW direction,
and $\mathcal{N}$ indicates the total number of lattice sites in the whole system.
Here, \(s\) is the DW sublattice index. Note that we construct the whole lattice by placing DW supercells along the DW, where atoms with similar positions in the supercells have the same \(s\), while different $x_i$.}
	{The spectral function gives the squared overlap amplitude between the numerically obtained eigenstates and the DW Bloch states at given energy $\omega$ and momentum $k$. }
	{The spectral function calculations have already been utilized in various contexts such as disordered systems and quasicrystals to study topological edge modes in the absence of translational symmetry ~\cite{PhysRevB.85.201105,PhysRevB.88.045429,PhysRevB.91.014202,PhysRevB.93.035123,PhysRevB.100.014510,PhysRevB.104.144511}.}

	{In Fig.~\ref{fig2B}, we study the influence of random onsite potentials with the strength $V_i\in [-W, W]$ on a finite-size system composed of around 285 supercells. Here, each supercell contains approximately $2 \times 162$ atoms.
	Disorder potential is introduced in the region of width 8b near the DW as in Fig.~\ref{fig1B}(a) which contains $2 \times 21$ atoms.  }

	{To check the validity of the spectral function method, in Fig.~\ref{fig2B} (a), we computed the spectral function of a disorder-free system and confirmed the same energy dispersion of helical DW states as in the periodic case shown in Fig.~\ref{fig1B}(b). }

	{In Fig.~\ref{fig2B} (b-e), we introduce disorder with the same configuration but increasing the strength of $V_i$. 
	We found that the helical DW states remain gapless even in the presence of large disorder strength  $W\approx 1$ eV [here, we choose $m=0.5$ eV].  
	As the disorder strength increases, the spectral function of DW states loses its spectral weight and sharp spectral features.}

	{Note that the disorder does not strongly affect the bulk modes (the states with $|E| \gtrapprox 0.5$ eV in Fig.~\ref{fig2B}) or edge mode (the in-gap state near E$\approx$ 0.4 eV in Fig.~\ref{fig2B}), but only affects the DW modes, because we considered disorder only in the region near the DW.}

	{In the presence of strong disorder with amplitude $W>1$ eV, one can observe a significant reduction in the amplitude of the spectral function (see  Fig.~\ref{fig2B} (f-i)). For instance, the spectral function loses its amplitude almost $80\%$ for $W=3$ eV. 
	This indicates that the spectral function loses its Bloch character under larger disorder due to scattering. 
	The stability of helical DW modes against disorder is aligned with previous studies showing that the topological edge modes can survive under disorder with a strength comparable to the bandwidth ~\cite{PhysRevB.85.201105,PhysRevB.88.045429}.}

	{According to the calculation carried out in Fig.~\ref{fig2B}, we believe that even with moderate disorder, ZBC QVHE helical modes remain stable, and thus are experimentally observable.}

	\begin{figure}[h!]
		\centering
		\includegraphics[width=\textwidth]{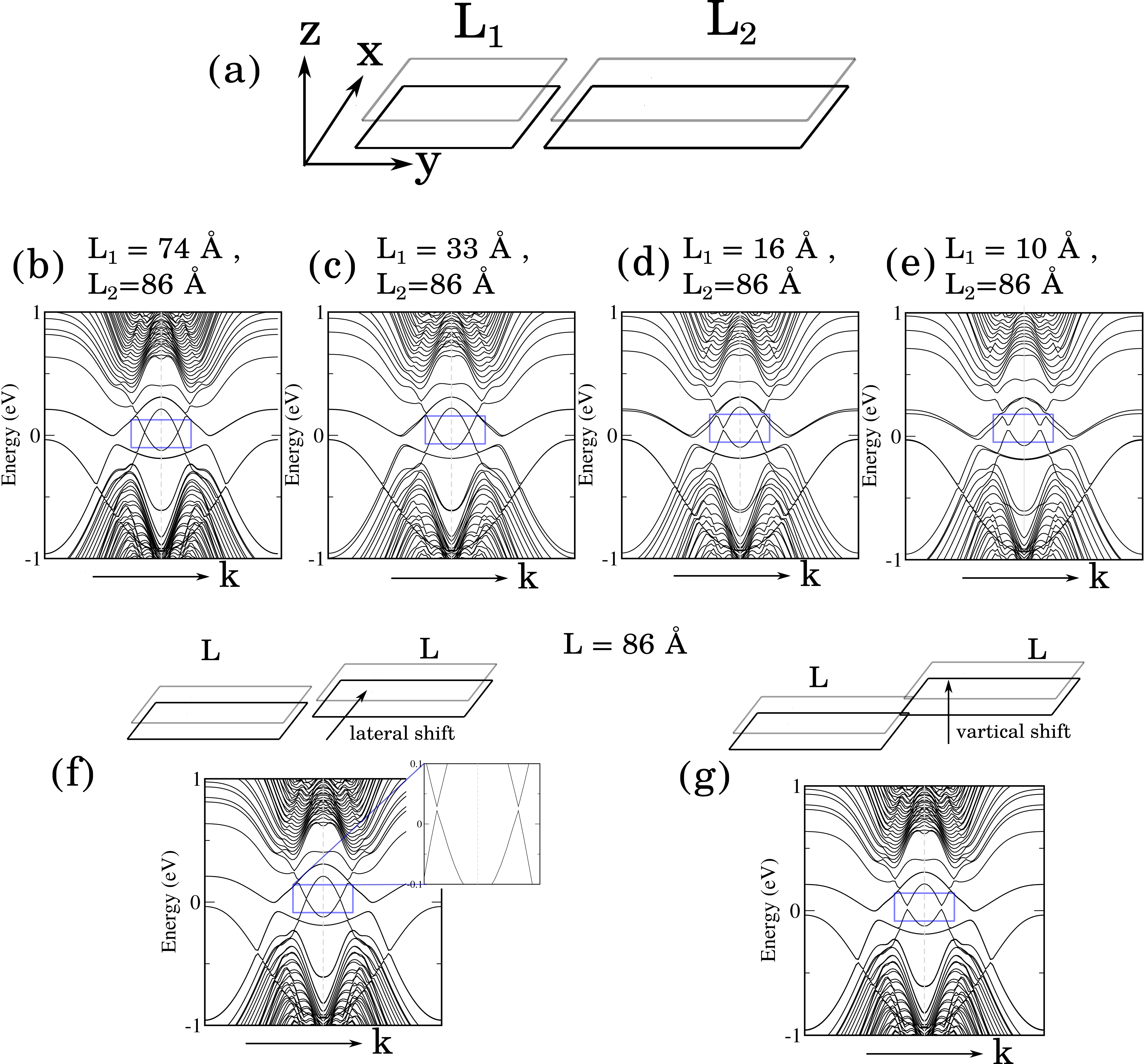}
		\caption{ 
			{DW mirror symmetry breaking effect on the stability of helical DW states. 
				(a) Schematic figure for DW mirror symmetry breaking due to unequal domain length ($L_1 \neq L_2$). 
				(b-e) The band structures for different $L_1$ while $L_2$ is kept fixed at 86 $\AA$.  
				Although the mirror symmetry is broken globally, as the atomic configurations near the DW are unchanged, the DW modes remain gapless when $L_1$ is large enough. 
				Small gap opening can be observed when $L_1$ becomes smaller than $16\AA$ (d,e).
				(f) The DW mirror breaking by laterally shifting the right side domain. The corresponding band structure shows slightly gapped domain wall modes (see inset). 
				(g) The mirror symmetry breaking by vertically shifting the right domain.}
		}
		\label{fig3B}
	\end{figure}
 
	{Additionally, we also investigate the effect of DW mirror symmetry breaking in the h-BAs system using DFT band structure calculations (see Fig.~\ref{fig3B}). 
	We consider three types of mirror-breaking in h-BAs systems; (i) breaking the DW mirror by taking unequal left and right side domain lengths [see Fig.~\ref{fig3B}(a-e)], (ii) rigidly shifting one side of the domain laterally [see Fig.~\ref{fig3B}(f)], (iii) rigidly shifting one side of the domain vertically [see Fig.~\ref{fig3B}(g)]. }

	{For unequal domain size on the two sides of the system, we consider the different lengths of the left domain $L_1$ and keep the right domain length fixed at $L_2 \approx 86 \AA $ to compute the band structures of the systems as shown in Fig.~\ref{fig3B}(b,e). The band structures of the systems with decreasing $L_1$ are shown in Fig.~\ref{fig3B}(b,e). In Fig.~\ref{fig3B}(b,c), the DW modes do not open up the gaps even though the mirror symmetry is broken. 
	However, the DW modes open up the gap when $L_1$ is further reduced as shown in Fig.~\ref{fig3B}(d,e). 
	This is because the electrons at DW and the left edges get hybridized for small $L_1$. 
	Although this length difference breaks the DW mirror symmetry globally, it does not significantly affect the helical DW modes for large domain lengths as the protecting DW mirror remains intact around the DW.
	This type of DW mirror breaking could often be present in experiments. However, we find that this type of mirror breaking does not affect the stability of the helical domain wall modes as long as the size of the two domains is large enough.}

	{We also study the DW mirror symmetry breaking by rigidly shifting one side of the domain either laterally or vertically (Fig.~\ref{fig3B}(f,g)). 
	In our simulations, we consider around 2.5 $\%$ of lateral shifting with respect to the optimized bond length between B and As atoms in h-BAs. 
	The optimized bond length in h-BAs is calculated to be 1.95 $\AA$.
	In the case of vertical shifting, we shift one domain vertically by 2.4$\%$ of the vertical spacing between the two layers which is 4.15 $\AA$.  
	Note that the amount of perturbation is significantly high in two cases.
	In both cases, the helical domain wall modes open up small gaps (as shown in the inset) but their overall spectrum remains robust [see Fig.~\ref{fig3B}(f,g)].  
	As the helical DW modes remain present even with these high mirror-breaking perturbations, we believe that under careful experimental measurements with the perturbations minimized, the properties of helical domain wall modes can be clearly measured.}

	{Considering our additional study of various perturbations illustrated in Fig.~\ref{fig1B}, Fig.~\ref{fig2B}, and Fig.~\ref{fig3B}, we can conclude that the helical DW states of ZBC QVHE are robust enough and their physics consequence could be observed experimentally by carefully designing the DW configuration.}

\clearpage
\newpage
  \bibliography{refs}